# Merger signatures in low excitation radio galaxies

David Garofalo

Department of Physics, Kennesaw State University, Marietta GA 30060

Abstract

While no consensus governs our understanding of the origin of low redshift radio galaxies, the possibility that mergers may trigger accretion from hot cluster halo gas has spurred a recent search for such signatures. Evidence for mergers is at best tenuous, however, and even when found, generates more questions than answers. With scant evidence for minor mergers, some connection to major mergers is found in isolated environments but not where one would expect, i.e. in clusters. We provide an explanation for these recent results by Gordon et al (2019) on the relevance of major mergers in low excitation radio galaxies (LERGs) at low redshift. While LERGs are not the direct result of a merger, we describe how they form in clusters in only a few million years while that timescale is an order of magnitude longer in field environments. As a result of these different timescales, the average lifetime of a cluster LERG is estimated at an order of magnitude greater value than for field LERGs. Observing a LERG in the cluster environment, therefore, will tend to occur when greater time has passed since the major merger event that produced its high excitation radio galaxy ancestor, such that fewer signatures of that event remain visible. We provide simple estimates for the fraction of LERGs as a function of environment that are directly related to these timescales, obtaining a probability of about 7% that field LERGs will show merger signatures and 3% for clusters, showing that theory and observation match if major merger signatures remain visible for a few hundred million years.

1. Introduction

Low excitation radio galaxies (LERGs) are thought to be accreting at low rates, with low radiative efficiency and with an absence of strong emission lines (e.g. Best & Heckman 2012). They prefer to live in cluster environments (Ramos-Almeida et al 2013; Hardcastle 2018) and the absence of merger signatures for LERGs is strong (Baum, Heckmann & van Breugel 1992; Hardcastle et al 2007; Baldi & Capetti 2009; Chiaberge, Capetti & Celotti 1999; Emonts et al 2010). Gordon et al (2019) have explored the connection between LERGs and both minor and major mergers as a function of environment and find that although neither dominates in LERGs, there is a statistically meaningful difference between major merger signatures observed in LERGs that live in field environments compared to denser, cluster regions. Mergers, therefore, appear to not only be weak triggers of LERGs, but to have a puzzling connection in their preference for field LERGs. So what triggers LERGs? There seems to be a contradiction in our assumption that the most powerful active galaxies are triggered in mergers. This is because the more powerful LERGs prefer cluster environments over isolated ones (Magliocchetti et al 2018). In short, the explanation for the formation of a LERG needs to include a weak and counterintuitive dependence on major mergers, with field LERGs experiencing a greater fraction of such signatures, while also explaining the emergence of more powerful LERGs in clusters.

In this paper we provide a context for understanding the origin and evolution of LERGs in both field and cluster environments, including redshift dependence, in a way that allows us to make quantitative contact with observations. We will show that because LERGs in general constitute later stages in the evolution of powerful radio galaxies, they are not directly associated with mergers. Despite this, their ancestors were radio quasars that on average formed in the more distant past in cluster environments compared to field environments which explains the greater visibility of merger signatures in field galaxies. We provide quantitative timescales showing that theory and observation match if major merger signatures can survive a timescale of a few hundred million years. In Section 2a we quantitatively describe the observations of Gordon et al (2019) and in Section 2b the applied model. In Section 3 we conclude.

2. a. LERGs v mergers: observational picture

Gordon et al (2019) studied the presence of minor and major merger signatures in a sample of 282 LERGs in both isolated and over dense environments at redshift below 0.07. While minor mergers have no distinguishable effect in isolated LERGs, their underrepresentation compared to the control sample suggests they may have an effect in cluster environments. Our interest in this work is primarily focused on their findings concerning major mergers which although weakly correlated with LERGs in general, show a statistically significant prevalence for LERGs in isolated environments compared to cluster LERGs. These results are shown in Figure 1 with field LERGs on the left and cluster LERGs on the right. While the fraction of merger signatures in cluster LERGs is smaller than in the control group, the opposite is true for field LERGs where major merger signatures not only dominate the control group but are also about twice that of the cluster LERGs. In the next section we derive similar numbers from a model that is founded on the idea that neither minor nor major mergers trigger LERGs.

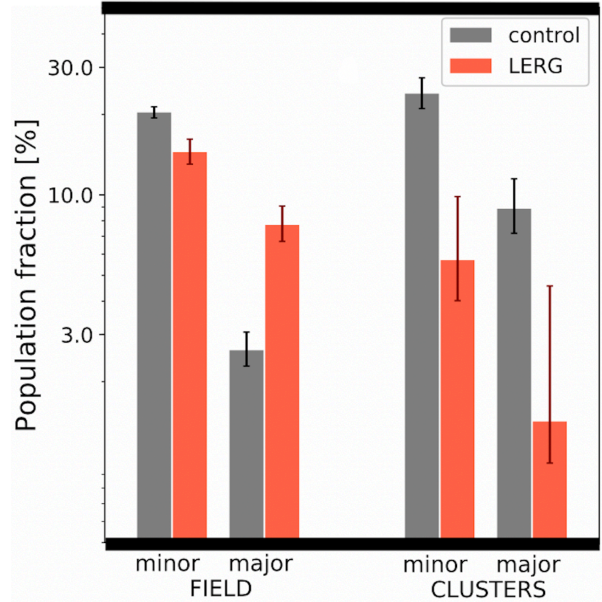

Figure 1: Results from Gordon et al (2019) showing about twice the population fraction for visible major merger signatures in field LERGs compared to cluster LERGs. We have extracted the values from Figure 6 of Gordon et al (2019) excluding the data for groups.

2. b. Origin and evolution of LERGs: theoretical picture

The gap paradigm for black hole accretion and jet formation (Garofalo, Evans & Sambruna 2010) purports to explain the active galaxy phenomena by combining into one framework ideas associated with jet formation (Blandford & Znajek 1977 and Blandford & Payne 1982), radiative disk winds (Kuncic & Bicknell 2004), and flux-trapping via the Reynolds condition (Reynolds et al 2006). Most recently the model has been applied to explain the inverse relation between winds and jets found in Mehdipour & Costantini 2019 (Garofalo 2019), the nature of the recently discovered FR0 radio galaxies (Garofalo & Singh 2019), the distribution in redshift of the blazar subset of powerful jetted AGN (Garofalo et al 2019), and the jet power/lifetime correlation and environmental dependence of radio quasars and FRI radio galaxies (Garofalo, Singh & Zack 2018).

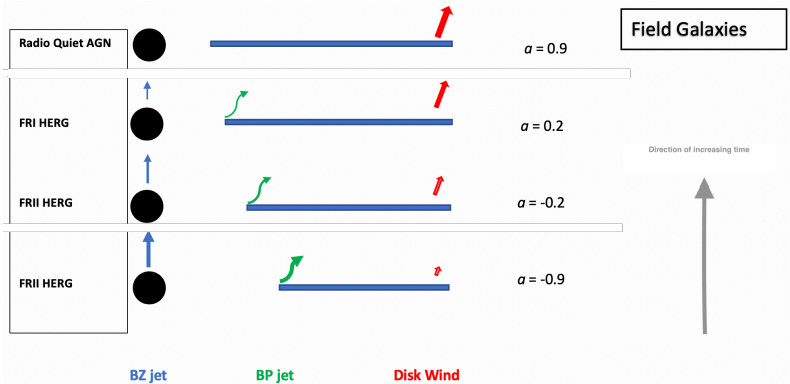

Figure 2: Time evolution of post-merger, counterrotating black holes (lowest panel). BZ represents the strength of the Blandford-Znajek jet, BP that of the Blandford-Payne effect, and the red arrows show the power in the radiative wind from the disk. Black hole spin is indicated on the right with negative values representing counterrotation and positive values co-rotation. Radio morphology and excitation class indicated on the left. HERG is the acronym for high excitation radio galaxy. Because of smaller dark matter halos in field environments, these smaller black holes produce weaker jet power. Gray arrow indicates the direction of time which increases upwards in the diagram as the spin goes from retrograde to prograde.

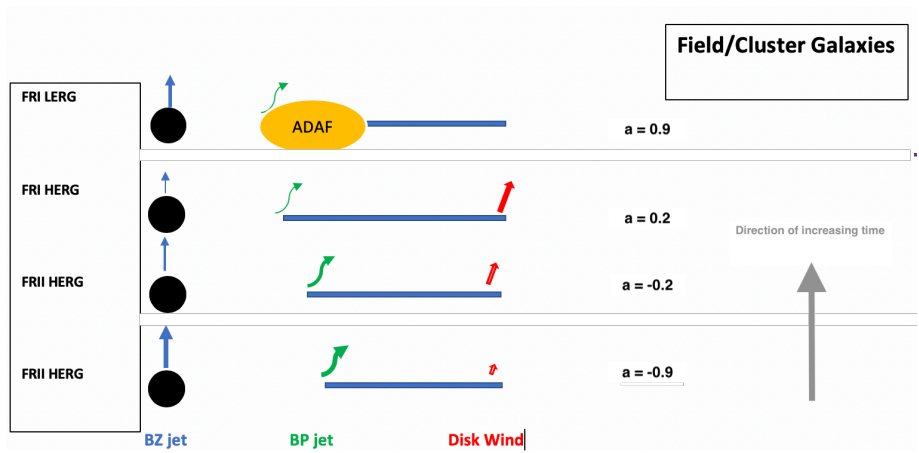

Figure 3: Same as in Figure 2 except for larger mass black holes that are distributed in both isolated as well as cluster environments. Since their black hole masses are on average larger than in Figure 2, their average feedback is greater and their accretion disks eventually evolve into ADAFs. LERG stands for low excitation radio galaxy.

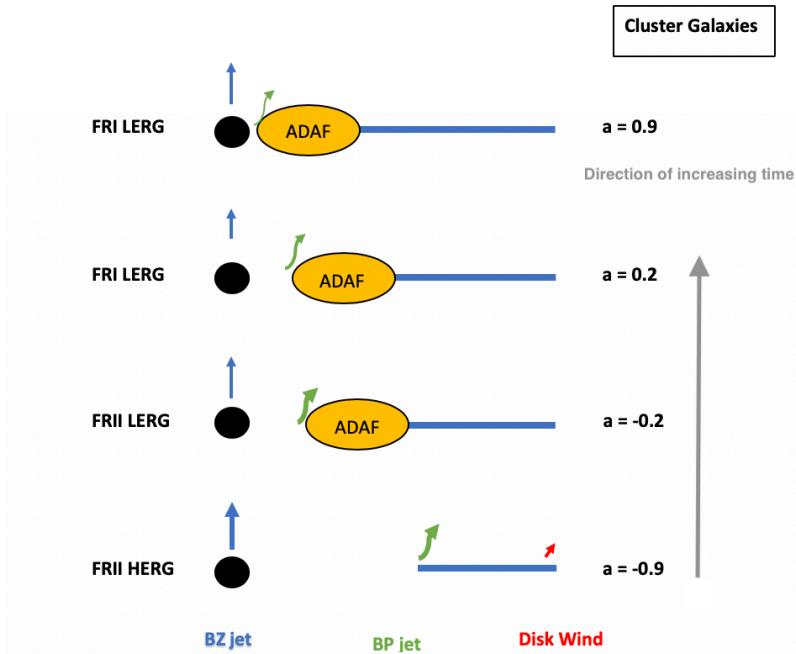

Figure 4: Same as in Figures 2 and 3 except for largest black holes that dominate in cluster environments. Because of their greatest average feedback, such systems evolve rapidly away from their original thin disk configurations while still in retrograde mode.

The basic idea can be illustrated using the cartoons of Figures 2,3, and 4. The model is anchored to the possibility that post-merger cold gas funneled into the black hole sphere of influence may, as a result of the Bardeen-Peterson effect, end up forming either a counterrotating or corotating accretion disk with respect to the rotating black hole. For systems where the black hole mass is dominant over the accreting mass, counterrotation is stable and, if it occurs, an FRII HERG or FRII quasar is formed. Because jet power in the model is a combination of the Blandford-Znajek effect (BZ) and Blandford-Payne mechanism (BP), not only does the overall jet power depend on the spin value of the black hole, it is also proportional to the square of black hole mass (Details for how the BP mechanism depends on black hole spin can be found in Garofalo, Evans & Sambruna 2010). Everything else being equal, we simplify the possible initial feedback of the newly triggered counterrotating black hole-accretion disk system by imagining three classes of black hole mass: below a few tens of millions of solar masses, between a few tens of millions to a few hundred million solar masses, and the most massive above about 500 million solar masses, as captured in Figures 2,3 and 4, respectively. Effective jet feedback will affect the ISM and eventually the accretion state (see Garofalo, Evans & Sambruna 2010 for details), which explains why the ADAF state for the disk appears on different panels, i.e. at different times in their time evolution.

Figures 2,3, and 4 describe the same initial conditions, namely cold mode accretion in a retrograde accretion disk, instantiating the conditions that produce powerful, collimated jets. However, for systems that live in more isolated or field environments with smaller dark matter halos, the black holes tend to be smaller and their jet feedback correspondingly weaker (Figure 2). Such systems fail to effectively heat their ISM, do less in the way of star suppression, and thus remain in cold accretion mode. As the spin evolves toward the high prograde state, the conditions that suppress jet formation come into play and the jet shuts off (top panel in Figure 2). Such systems become radio quiet quasars or radio quiet AGN. For larger black holes whose jet feedback is more effective (Figure 3), the ISM is heated, star formation eventually shuts down, and accretion transitions from cold mode to advection dominated mode (ADAF). But this process requires on the order of a few tens of millions of years by which time the black hole and accretion disk are in co-rotation. This timescale is not arbitrary but comes from estimating the growth of the black hole as a result of its feeding from the thin accretion disk at the Eddington accretion limit (Moderski et al 1998; Kim et al 2016).  Because these black holes are on average more massive, they live in either field or in cluster environments (Figure 3). For the largest black holes whose jet feedback is most effective (Figure 4), the ISM is heated rapidly, star formation quickly shuts down, and accretion transitions to ADAF mode in less than 10 million years. This timescale also emerges from Eddington-limited accretion in that it takes 8 million years to completely spin down a rapidly spinning black hole (Moderski, Sikora & Lasota 1998; Kim et al 2016) and since the ADAF state sets in when the system is only about halfway to becoming a Schwarzschild black hole, it will take only about 4 million years for that to occur. Because ADAF mode accretion slowly builds the black hole, such systems evolve slowly and can last billions of years as the large reservoir of gas is accreted at low rates. Note that ADAF accretion is associated with low excitation, hence the term LERG.

We now show, from the ideas captured in Figures 2, 3, and 4, how to extract the probability that an observation of a LERG occurs within the merger signature relaxation time, i.e. the chance that the observation of the LERG is made during the time when merger signatures are still visible. Despite the fact that LERGs weakly inhabit field galaxies as shown in Figures 2 and 3 (see more in-depth analysis of this in Garofalo, Singh & Zack 2018), these figures also allow us to understand why field LERGs are more likely to display signatures of major mergers as we now show. As mentioned, Eddington-limited accretion will spin down a maximally spinning black hole via counterrotation in about $8 \times 10^6$ years and up to maximum spin again via a co-rotating disk in about another $10^8$ years.  Because Figure 3 shows an ADAF forming only after the accretion disk has reached the prograde configuration, a non-negligible fraction of $10^8$ years must have passed prior to it displaying signs of being a LERG. Compare this with the behavior in clusters (Figure 4). There, ADAFs appear early while the disk is still in counterrotation. As a result, field galaxy LERGs on average accrete more of their black hole mass from disks near the Eddington limit compared to those in cluster galaxies. Therefore, cluster black holes on average live more of their lives as LERGs than their field galaxy counterparts.

We can be precise about the consequences of this by assuming for simplicity that accreting black holes on average have a characteristic amount of mass, enough fuel to spin the black hole up to maximum prograde spin values. This means there is a supply of gas that is on the order of twice the mass of the original black hole (Figure 5). For black holes in field galaxies that transition to LERGs (Figure 3), the LERG state occurs when the mass has increased to about 1.5 times the original black hole mass (Figure 5). This means there is still at least 0.5 times the original black hole mass available to accrete, which would take about 50 million years at the Eddington limit but as an ADAF it will take at least 5 billion years because ADAFs set in at accretion rates below about 1% that of Eddington. For the most massive black holes, on the other hand, they have been in ADAF states since the spin was intermediate and the orientation was still retrograde. Hence, the time needed to spin such black holes up to high prograde values is on the order of at least 10 billion years. In short, the timescale for life as an ADAF is twice as long for the more massive black holes.

With these timescales in hand we are close to being able to answer the following question: What is the probability that an observation of a LERG will occur when the LERG is close enough to the major merger that caused the FRII HERG state such that merger signatures are still visible? While mergers take hundreds of millions of years, delays between mergers and AGN ignition needs to be accounted for. Because field galaxy LERGs live less in ADAF states compared to cluster galaxies (since much of their fuel is accreted in thin disks), the probability of observing them closer in time to active nucleus ignition is greater. We can be quantitative here. The probability of catching the LERG close enough to see merger signatures will depend on the relaxation time for merger signatures, the time that passes between the merger and the onset of the LERG state, as well as the time the system lives as a LERG, with that probability inversely proportional to the LERG lifetime. If merger signatures last throughout the LERG lifetime, that probability is 100%. As long as the LERG phase dominates, the fractional probability can be constructed from the timescales discussed above as follows.

$$P = (T_{ms} - T_{pre\text{-}LERG})/ T_{ADAF}$$

where $T_{ms}$ is the timescale during which merger signatures remain visible, $T_{pre\text{-}LERG}$ is the time that passes between the merger and the beginning of the ADAF phase, and $T_{ADAF}$ is the lifetime of the ADAF or LERG. For field galaxies the ADAF phase occurs at some point in the prograde configuration so $T_{pre\text{-}LERG}$ includes a time associated with the HERG state, $T_{HERG}$, which amounts to a few tens of millions of years for the system to reach the prograde regime, say $5 \times 10^7$ years, plus any difference in time between the merger and AGN triggering. If ~75% of the cold gas turns into stars on timescales of about 75 million years after a merger (Somerville et al 2001) coupled with the idea that AGNs turn on while stars are actively forming, one might estimate a characteristic time difference between merger and AGN triggering of 75 million years. In minor mergers, however, that delay may be up to an order of magnitude greater (Shabala et al 2017). Although we are focused on major mergers, these results suggest large uncertainties. Merger signatures last ~ 1 billion years (Lotz et al 2008; Ji et al 2014) so $T_{ms} - T_{pre\text{-}LERG}$ is likely to be at least a few hundred million years. In order for observation to match theory, we choose $T_{ms} - T_{pre\text{-}LERG}$ to be 350 million years. Because spinning the black hole up occurs in the ADAF phase which takes at least two orders of magnitude longer than at the Eddington accretion rate, we estimate $T_{ADAF}$ ~ $10^2 \times 5 \times 10^7$ years = $5 \times 10^9$ years. For the average field LERG, therefore, the fractional probability of observing a merger signature is

$$P_{field} = (3.5 \times 10^8)/(5 \times 10^9) = 0.07.$$

For the average cluster LERG, on the other hand, $T_{HERG}$ is some fraction of the retrograde spin down phase, which is less than the $T_{HERG}$ for fields, say $4 \times 10^6$ years (the entire retrograde spin down phase is about $8 \times 10^6$ years at the Eddington rate). This comes from Figure 4 where the ADAF or LERG state is reached when the system is only halfway through its retrograde phase. Hence, to reach maximum prograde spin, the system needs to spin down to zero and up again in ADAF mode, which slows it down by a factor of at least 100. As in fields, the $T_{HERG} = 4 \times 10^6$ years in clusters contributes little to changing $T_{pre-LERG}$ while $T_{ADAF} \sim 100(4 \times 10^6 + 10^8)$ years. This gives a merger signature probability in the average cluster LERG of

$$P_{cluster} = (3.5 \times 10^8)/(4 \times 10^8 + 10^{10}) = 0.033.$$

The difference in the probabilities is contingent only on $T_{ADAF}$. In percentages, we find about 7% for field LERGs and about 3% for cluster LERGs. Our $T_{ADAF}$ values are also uncertain and could in fact be longer by orders of magnitude. If so, however, $T_{ADAF}$ becomes longer than the age of the Universe and simply gets replaced by the latter time. Hence, our estimate for the probabilities must satisfy $T_{ADAF} < T_U$. If we allow $T_{ADAF}$ to be as large as $T_U$, we end up with $\Delta P = 0.045$ in fields and $\Delta P = 0.0086$ in clusters. These simple estimates are surprisingly compatible with the results of Gordon et al (Figure 1).

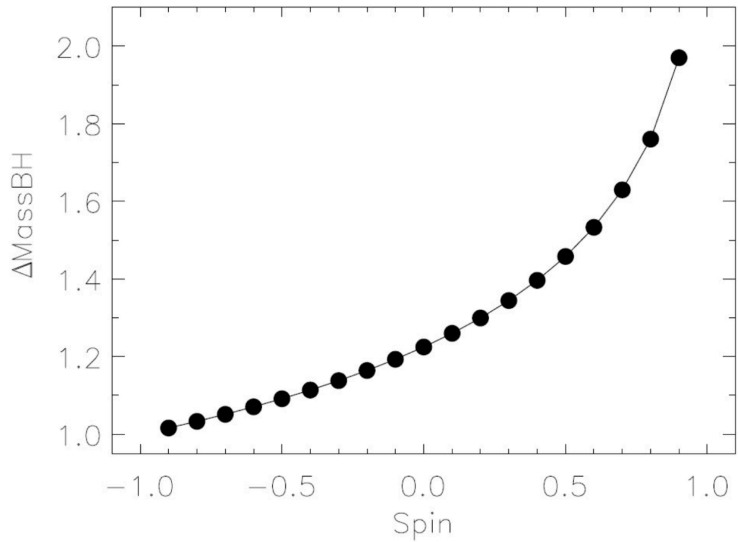

Figure 5: Increase in black hole mass necessary to spin the black hole down from counterrotation into co-rotation (from Figure 1 of Kim et al 2016).

3. Conclusions

Because FRII quasars with weaker jet feedback tend to live in field environments due to their less massive black holes, such systems rapidly spin up toward prograde configurations but evolve slowly if at all in their accretion properties, in heating their ISMs, and in suppressing star formation. This means that any tendency to form a LERG in field environments occurs when more of the accretion fuel is consumed such that LERG lifetimes are shorter on average in fields compared to clusters. In contrast to fields, LERG states in clusters are reached early which means these environments will tend to have more LERGs that dominate the lifetime of the active galaxy compared to active galaxies in fields. Somewhat counterintuitively, note that cluster LERGs come into play at early times, when merger signatures are strongest, whereas field LERGs form about an order of magnitude later times after the merger event that triggered the FRII quasar. These $T_{HERG}$ times are, however, small compared to the dominant $T_{ADAF}$ which is mostly responsible for washing away the merger signs in clusters. Note also that in our model a natural redshift dependence among HERGs and LERGs emerges from the accretion process itself which by its nature enforces prograde accretion states to live in the future of retrograde accretion configurations. Along those same lines, we see that in the model LERGs are not young objects and cannot evolve into HERGs.

Acknowledgments

I thank the anonymous referee for a detailed report that led to improvement in the figures and to a more meaningful analysis of the relevant timescales.